\newcommand{\beque}{\begin{equation*}}
\newcommand{\eeq}{\end{equation}}
\newcommand{\beq}{\begin{equation}}
\newcommand{\eeque}{\end{equation*}}
\newcommand{\beqnl}{\begin{eqnarray}}
\newcommand{\eeqna}{\end{eqnarray*}}
\newcommand{\beqna}{\begin{eqnarray*}}
\newcommand{\eeqnl}{\end{eqnarray}}

\documentclass[aps,prl,twocolumn]{revtex4}

\usepackage{amsfonts}
\usepackage{amsthm}
\usepackage{amsbsy}
\usepackage{amssymb}
\usepackage{amsmath}
\usepackage{graphicx}
\usepackage{color}

\begin{document}
\date{\today}

\title{Negative frequency resonant radiation.}
\author{E. Rubino,$^{1,*}$ J. McLenaghan,$^{2,*}$ S. C. Kehr,$^{2}$ F. Belgiorno,$^{3}$, D. Townsend,$^{4}$ S. Rohr,$^{2}$ C.E. Kuklewicz,$^{4}$  U. Leonhardt,$^{2,*}$   F. K\"onig,$^{2,*}$  and D. Faccio$^{1,4,*}$ }
\address{$^{1}$ Dipartimento di Scienza e Alta Tecnologia, Universit\`a dell’Insubria, Via Valleggio 11, IT-22100 Como, Italy\\
$^{2}$ School of Physics and Astronomy, SUPA, University of St Andrews, North Haugh, St Andrews, KY16 9SS, UK\\
$^{3}$  Dipartimento di Matematica, Politecnico di Milano, Piazza Leonardo 32,20133 Milano, Italy\\
$^{4}$ School of Engineering and Physical Sciences, SUPA, Heriot-Watt University,
Edinburgh EH14 4AS, UK
}

\email{D. Faccio: d.faccio@hw.ac.uk; F. Konig: fewk@st-andrews.ac.uk}

\begin{abstract}
Optical solitons or soliton-like states shed light to blue-shifted frequencies through a resonant emission process. We predict a mechanism by which a second propagating mode is generated. This mode, called negative resonant radiation originates from the coupling of the soliton mode to the negative frequency branch of the dispersion relation.  Measurements in both bulk media and photonic crystal fibres confirm our predictions.
\end{abstract}


\maketitle
{\emph{Introduction - }}
 Resonant radiation (RR), often also referred to as dispersive-wave or Cherenkov radiation is a nonlinear optical process by which a soliton propagating in an optical fibre in the presence of higher-order dispersion sheds light through a resonant-like process to a shifted frequency \cite{Wai,cerenkov,cristiani,Skryabin,dudley}. This process and the precise frequency of the  RR is determined by a wave-vector conservation relation, 
 \begin{equation} \label{RR}
 k(\omega_\textrm{RR})-k{\omega_\textrm{IN}}-(\omega_\textrm{RR}-\omega_\textrm{IN})/v-K_\textrm{NL}=0
 \end{equation}
  where $\omega_\textrm{IN}$ and $\omega_\textrm{RR}$ are the soliton and  RR frequencies, $v$ is the soliton velocity and $K_\textrm{NL}=\omega_\textrm{IN}n_2I/c$ is a nonlinear correction term that may be small or even negligible at low intensities, $I$ \cite{dudley}. A very similar process occurs also in bulk media. The stationary 1D fibre soliton is now replaced by the stationary 3-dimensional X-wave \cite{recami}. X-waves may form spontaneously in Kerr media at high enough powers in much the same way that solitons form spontaneously in a fibre \cite{miroPRL,faccioPRL}. A blue-shifted peak will also be observed that will form one of the two X-wave tails: the whole X-wave, including the  RR is therefore described by Eq.~(\ref{RR}) \cite{faccio}, which indeed reflects the non-dispersive nature of the wave-packet considered, i.e. the soliton in 1D and the X-wave in 3D. \\
  These frequency conversion processes may be understood in terms of energy transfer between specific modes identified by Eq.~\eqref{RR} and the dispersion curve \cite{Skryabin,dudley,faccio}. However, to date only the positive frequency branch of the dispersion has been considered when this actually also has a branch at negative frequencies. This branch is usually neglected or even considered meaningless when, in reality, it may host mode conversion to a new frequency.  The fact that a mode on the negative branch of the dispersion relation may be excited has a number of important implications, beyond the simple curiosity of the effect in itself. Indeed, light always oscillates with both positive and negative frequencies,
but the negative-frequency part is directly related to its positive counterpart and seems redundant \cite{born}. On the other hand, light particles, photons, have positive energies and are associated with positive frequencies only \cite{mandel}. A process such as that highlighted here, that mixes positive and negative frequencies will therefore change the number of photons, leading to amplification or even particle creation from the quantum vacuum \cite{birrell,brout}. \\
In this work we show how alongside the usual resonant radiation spectral peak observed in many experiments, a second, further blue-shifted peak is also predicted. This new peak may be explained as the result of the excitation of radiation that lies on the negative frequency branch of the dispersion relation.  We first explain why this radiation should be observed and then provide experimental evidence of what we call ``negative frequency resonant radiation'' in both bulk media and photonic crystal fibres.\\
\begin{figure}[t]
\centering
\includegraphics[width=7cm]{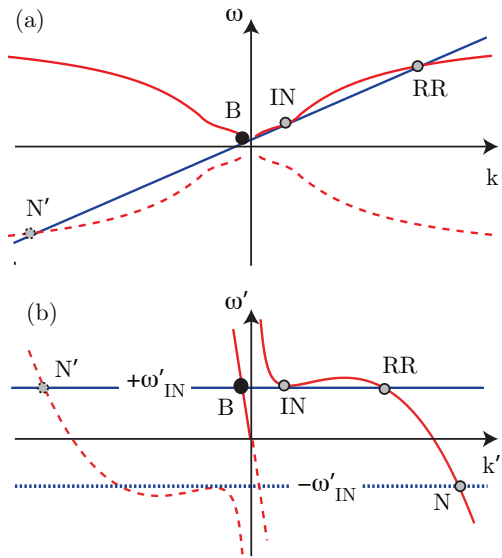}
\caption{Typical dispersion relation, e.g. for fused silica glass, in the laboratory reference frame (a) and in the reference frame comoving at the soliton velocity (b).  Dashed curves indicate the (laboratory frame) negative frequency branches of the dispersion relation.}
\label{fig1}
\end{figure}
{\emph{Theory - }} In order to show how the negative frequency  RR arises, we consider without any loss of generality a basic dispersion relation that contains higher order terms such as in fused silica glass and shown in Fig.~\ref{fig1}(a). The dashed curves indicate the negative frequency branches. The phase-matching relation \eqref{RR} (with $k_\textrm{NL}=0$) is a straight line that intersects the dispersion relation in a number of points that define the allowed modes. The point indicated with IN is simply the input mode, or soliton mode, and B indicates a backward propagating mode that is not excited in the experiments and is therefore neglected. There is then a third positive frequency mode, RR, that indicates the resonant radiation  mode. However, there is also a negative frequency mode, N', equally predicted by Eq.~(\ref{RR}) yet always neglected. The object of this work is precisely this negative frequency branch mode. All these various modes are easier to visualise in the comoving reference frame coordinates as shown in Fig.~\ref{fig1}(b). These curves are obtained from the original dispersion relation by transforming to the comoving coordinate system $\omega'=\gamma(\omega-vk)$ and $k'=\gamma(k-\omega v/c^2)$, with $\gamma=1/\sqrt{1-v^2/c^2}$ and $v=v_g$ is the soliton group velocity. Transforming also Eq.~(\ref{RR}) to the comoving frame using the same relations, gives $\omega'=\omega'_\textrm{IN}$. In other words, momentum conservation in the laboratory reference frame corresponds to energy conservation in the comoving reference frame. The allowed modes are therefore now found by simply tracing a horizontal line through the input soliton mode (that by definition, has zero group velocity in the comoving frame and thus lies at a local minimum) and as before, looking for the intersections with the dispersion relation. The main point here is that the dispersion curves tell us that it should be possible, starting from two positive modes, IN and RR, to excite a third negative, N', mode. When trying to assign a physical meaning to the negative frequency mode we should recall that in reality any electromagnetic field is a real-valued quantity that can be written as a sum of a complex term with its complex coniugate: $E\sim\cos\omega t=\exp(+i\omega t)+\exp(-i\omega t)$. However, considering only the modes obtained from the intersections with $\omega'=+\omega'_\textrm{IN}$,  amounts to considering only the first complex term and neglecting the complex coniugate. In order to recover the full field we obviously need to also sum the modes obtained from the intersections with $\omega'=-\omega'_\textrm{IN}$: the sum of N' and N in Fig.~\ref{fig1}(b) therefore will give a real-valued field with a positive frequency in the laboratory reference frame. Nevertheless, as explained above, the origin of this mode lies in the coupling of one or more modes on the {\emph{positive}} frequency branch of the dispersion relation to a mode that lies on the {\emph{negative}} frequency branch. We also note that the negative mode has a truly distinct frequency from all the other modes in Fig.~\ref{fig1} and, if it is generated, it should appear as a clearly distinct peak in the spectrum with a higher frequency than the  RR mode. In analogy with the usual positive frequency  RR, we call this new mode ``negative frequency resonant radiation''.  \\
We note that the fact that the negative  RR mode is a solution to Eq.~(\ref{RR}) does not, alone, imply that it will actually be excited. The negative  RR mode lies on a dispersion curve that is separate from the positive mode branch of the IN and RR modes and it is  not guaranteed that energy may be transferred between the two. A similarity can be sought for example with higher-order spatial modes in a waveguide that also lie on separate dispersion curves. In the presence of an adiabatic (spatial) variation of the waveguide there will be no energy exchange between the various dispersion curves and higher order modes will not be excited.  Conversely, any sudden changes in the waveguide geometry will lead to an energy exchange between the modes. Similarly, in our situation any adiabatic or smooth (temporal) perturbation of the medium will not excite the negative  RR mode: a non-adiabatic variation within the input pump pulse, e.g. a steep shock front, is required. Indeed, preliminary numerical simulations albeit in simplified setting (see e.g. \cite{njp}) do indicate that the actual intensity of the negative mode depends critically on the steepness of the refractive index variation induced by the nonlinear Kerr effect.\\
Finally, we note that in the comoving frame both the  RR mode and the negative  RR mode propagate with negative group velocities (as can be deduced from the slope of the dispersion curve at these frequencies), i.e. in the backward direction. The phase velocities of the two modes are however opposite to each other. Conversely, in the lab. frame both the RR and the N modes have positive phase and group velocities, i.e. they both propagate in the forward direction. \\
{\emph{Experiments - }} We performed two sets of experiments in order to capture the formation of the negative  RR mode: (i) in a bulk medium  and (ii) in a few mm long photonic crystal fibre.
\\
\begin{figure}[t]
\centering
\includegraphics[width=8cm]{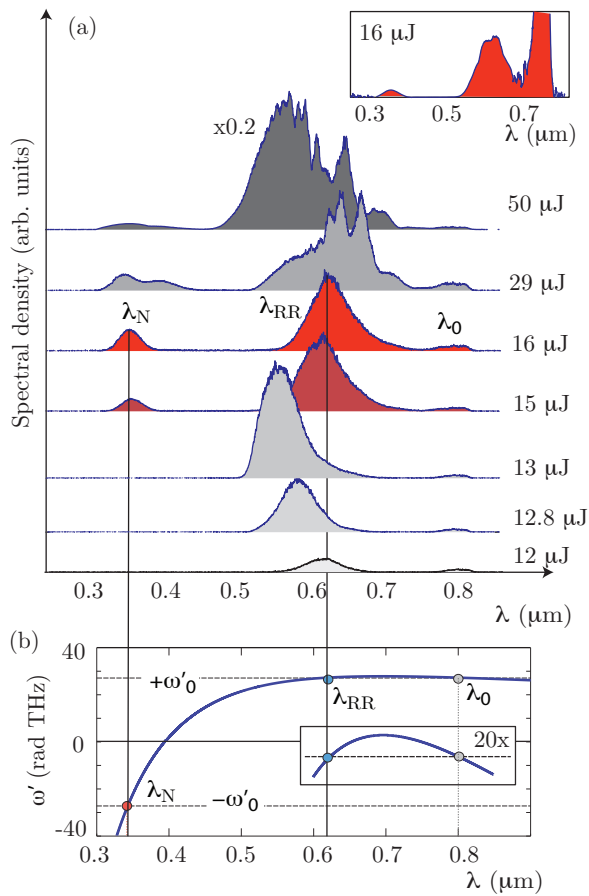}
\caption{Experimental results for negative  RR generation in bulk CaF$_2$. (a) shows the measured spectra for increasing input energies (indicated next to each curve). The spectra are vertically displaced to increase visibility. The inset shows a sample spectrum (16 $\mu$J input energy) corrected for the filter response. (b) shows the CaF$_2$ dispersive relation in the comoving frequency versus lab. frame  wavelength coordinates. Positions of the predicted  RR and negative  RR spectral peaks are indicated. The inset is a 20x enlargement of the curve around the $\lambda_\textrm{RR}$ wavelength.}
\label{fig2}
\end{figure}
In the first experiment, we chose a 2 cm long bulk calcium  flouride, CaF$_2$ sample as host material. Light pulses of 60 fs duration and 800 nm carrier wavelength
are provided by an amplified Ti:sapphire laser system of 1 kHz repetition rate. We reshape the pulses into Bessel
beams with a cone angle (in the medium) of $\theta=0.6$ deg, using a conical lens of fused silica with 2 deg base angle. The Bessel pulse in the sample moves with uniform speed  $v=v_g/\cos\theta$ where $v_g$ denotes the group velocity of a Gaussian pulse of carrier wavelength 800 nm. The spectrum at the output from the sample is collected with a lens and a fibre-based spectrometer. A  filter with a flat response in the visible-near-UV region  is placed before the spectrometer in order to reduce the input pump intensity without affecting the shape of the spectrum between 300 nm and 720 nm. The input pulse energy is varied from 10 $\mu$J  up to 50 $\mu$J at which point the input pulse is in a strongly nonlinear regime and develops a complex and structured spectrum. Generation of negative  RR modes is observed at intermediate energies $\sim15$ $\mu$J. Examples of the resulting spectra for varying input energies are shown in Fig.~\ref{fig2}(a). The spectra are vertically displaced  in order to render them visible. At lower energies (12-14 $\mu$J) the output spectrum shows a distinct single peak that shifts to shorter wavelengths with increasing input energy. This process has been described in detail \cite{blue-peak} in similar conditions and is a direct manifestation of the formation of a steep shock front on the trailing edge of the pump pulse. As energy is increased the shock front steepens and the spectral peak shifts towards shorter wavelengths. Between 15 and $\sim20$ $\mu$J input energy a different regime sets in, characterised by two distinct peaks in the spectrum that do not shift with increasing energy. The first peak is located around 620 nm, the second is much weaker and is located around 341 nm wavelength. Examples of these spectra (15 and 16 $\mu$J) are shaded in red in the figure. At higher input energies the pulse starts to develop complex dynamics, typical of the filamentation regime during which the pulse breaks up and creates a broad-band and highly structured spectrum known as white-light supercontinuum \cite{couairon}. We focus our attention for example on the spectrum measured for an input energy of 16 $\mu$J: the spectrum is not substantially modified if we account for the filter response, as shown in the inset to Fig.~\ref{fig2}(a). Three clear peaks are indicated with $\lambda_0$, $\lambda_\textrm{RR}$ and $\lambda_\textrm{N}$ and we indentify these with the IN, RR and negative RR modes, respectively. Indeed, these correspond exactly to the positions for the  RR and negative  RR modes given the IN mode and the dispersion relation for CaF$_2$ \cite{caf}, as shown in Fig.~\ref{fig2}(b).\\
We note that attempts to generate similar features in other glasses or media, e.g. BK7, fused silica and water, failed. Spectral broadening through self-phase modulation and the steepness of Kerr-induced shock fronts are both strongly limited by dispersion. Our experiments in fused silica and water (data not shown) showed that even at the highest input energies, spectral broadening exhibited a sharp cut-off around $\sim450$ nm, whereas the negative  RR peak was, in all cases, predicted to appear at shorter wavelengths. On the other hand CaF$_2$ (as other fluoride glasses) is quite unique as it exhibits significantly lower dispersion \cite{caf}, in particular in the UV spectral region and thus allows the formation of steeper shock fronts and broader continua. In this specific case it allows a relatively efficient excitation of the negative  RR peak in the UV.\\
\begin{figure}[t]
\centering
\includegraphics[width=8cm]{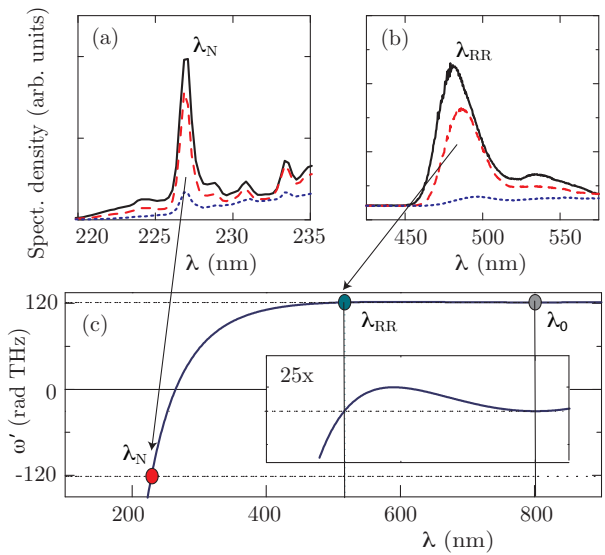}
\caption{Experimental results for negative  RR generation in a photonic crystal fibre. (a) and (b) show the measured spectra in the visible and UV regions for three different input energies.  (c) shows the fibre dispersive relation: positions of the predicted  RR and negative  RR spectral peaks are indicated. The inset is a 25x enlargement of the curve around the $\lambda_\textrm{RR}$ wavelength.}
\label{fig3}
\end{figure}
In a second experiment, we sent 7 fs light pulses, centred around 800 nm, with
a 77 MHz repetition rate into a fused silica photonic-crystal fibre. Photonic crystal fibres have the advantage of enhanced nonlinear effects due to tight mode confinement combined with a remarkable flexibility in tailoring the waveguide dispersion that can therefore strongly modify the corresponding bulk medium dispersion and thus allow observation and control of a variety of novel effects (see e.g. \cite{dudley} for an extensive review). We selected fibres where the spectrum of the
incident light lies in a region of anomalous group-velocity dispersion such that it
can propagate as a soliton-like pulse. 
The 7 fs input pulses are coupled into the fiber using a 90 deg off-axis parabolic mirror. We estimate the coupling efficiency to be 20\%. In a preliminary experiment we confirmed that most of the  RR emerged using only a few millimeters of  fiber. Therefore, short pieces of fiber of approximately 4-5 mm are used. Figure~\ref{fig3} shows the output spectrum after
5 mm of fibre for three different input energies (246, 324 and 366 pJ). The ultraviolet part of the spectrum, Fig.~\ref{fig3}(a) was measured with a monochromator and a photomultiplier tube and shows a clear peak that we identify with the negative  RR mode. The part of the output spectrum that lies in the visible range, shown in Fig.~\ref{fig3}(b), was
measured with a compact CCD spectrometer. The peak observed here corresponds to the  RR: the frequency of this mode shifts to shorter wavelengths with increased input pulse energy due to the nonlinear modification of the refractive index from the pulse \cite{dudley}. 
 Figure~\ref{fig1}(c) shows the predicted  RR and negative  RR
frequencies based on the dispersion relation for the photonic crystal fibre. The measured peaks at $\lambda_\textrm{RR}$ and $\lambda_\textrm{N}$ are, similarly to the bulk measurements, the main spectral features in the whole spectrum and both correspond very precisely to the predictions. We note that the negative  RR peak does not shift noticeably with input energy because the nonlinear refractive index change from the pulse is negligible compared to the dispersive index changes in the UV. \\
\begin{table}[t]
\begin{center}
\begin{tabular}{|l|c|c|c|}\hline
\rule[-3mm]{-1mm}{8mm}
Fibre& $\lambda_\textrm{RR}$&$\lambda_\textrm{N}$ pred.&$\lambda_\textrm{N}$ meas.\\[0.5ex]
\hline 
\rule[-3mm]{-1mm}{8mm}
NL-1.6 615a& 542 nm&233.4 nm&233.1 nm\\[-0.6ex]
NL-1.6 615b&  542 nm&233.3 nm&232.1 nm\\
NL-1.5 590       & 516 nm&228.7 nm&227.0 nm\\
NL-1.5 670a &478 nm&221.4 nm&218.1 nm\\
NL-1.5 670b& 480 nm&221.8 nm&218.9 nm\\ \hline
\end{tabular}
\caption{Predictions of $\lambda_\textrm{RR}$ and predictions and measured values for
$\lambda_\textrm{N}$ for the three fibres used in the experiment. \label{comparison}}
\end{center}
\end{table}
Experiments were repeated for a series of photonic crystal fibres as shown in Table~\ref{comparison}:  NL-1.6 615  and NL-1.5 590 (used for the data shown in Fig.~\ref{fig3}) consist of a solid silica core surrounded by a hexagonal pattern of air holes \cite{knight} and the fibre indicated with NL-1.5 670 consists of a solid silica core surrounded by a cobweb of silica strands. The size and spacing of these holes and the thickness ($\sim1$ $\mu$m) of the strands determine the dispersion profile of the fibre. The letters ``a'' and ``b'' in the fibre name indicate measurements performed along one of the two orthogonal polarization-maintaining axes of the fibre. The table lists the wavelength of the negative  RR emission predicted from the relative dispersion relations compared to the actually measured wavelengths: as can be seen, very good agreement is obtained in a variety of settings.\\
{\emph{Conclusion - }}
Frequency conversion through a resonant transfer of energy from an input laser pulse to a typically blue shifted peak is a well-studied process in nonlinear optics and has attracted substantial attention in the last few years due to the high conversion efficiencies that are attainable with short pulses \cite{cristiani,kartner} and more recently even to predicted mode-squeezing properties \cite{biancalana}. Here we have shown how the same process  generates a second, so far un-noticed peak that corresponds to resonant transfer of energy to the negative frequency branch of the dispersion relation. The energy transfer is favoured in the presence of steep shock fronts or more generically, by a non-adiabatic variation within the pump pulse. Experiments were performed in both bulk media and waveguides with optimised dispersion landscapes so as to allow the process to occur with a relatively high efficiency.  These experiments also conceptually demonstrate the classical limit of an important quantum effect, i.e. the particle creation by the mixing of positive and negative frequencies \cite{birrell}. This implies that the process could, if sufficiently optimised, generate a squeezed vacuum state that couples wavelengths around $\lambda_\textrm{RR}$ to wavelengths at the negative  RR peak, $\lambda_\textrm{N}$. At a classical level, the mixing of positive and negative frequencies leads to simultaneous wavelength conversion to the UV region and {\emph{amplification}} (at the UV wavelengths) of an input seed pulse that co-propagates with the input pump pulse. These ideas will be developed in future work.\\
We thank Simon Horsely, Thomas Philbin, Scott Robertson, Philip Russell and Sahar
Sahebdivan for discussions. ER wishes to acknowledge Fondazione Cariplo, Univercomo and Banca del Monte di Lombardia for financial support. Our work was supported by Heriot-Watt University, the University of St Andrews
and the Royal Society.\\

$^*$ E.R. and J.M. contributed equally to this work.\\

\end{document}